\documentclass[prb, showpacs]{revtex4}
\usepackage{graphicx}
\usepackage{dcolumn}
\usepackage{bm}
\graphicspath{{images/}}
\usepackage{epstopdf}

\begin{document}

\title
{Electromagnetic properties of a double layer graphene system with
electron-hole pairing}
\author{K.\,V.\,Germash$^1$, D.\,V.\,Fil$^{1,2}$}
\email{fil@isc.kharkov.ua}
\affiliation{$^1$Institute for Single Crystals, National Academy of
Sciences of Ukraine, Nauki ave. 60 Kharkov 61001, Ukraine\\
$^2$Karazin Kharkov National University, Svobody Sq. 4, Kharkov
61022, Ukraine}

\begin{abstract}
We study electromagnetic properties of a double layer graphene
system in which electrons from one layer are coupled with holes from
the other layer. The gauge invariant linear response functions are
obtained. The frequency dependences of the transmission, reflection
and absorption coefficients are computed. We predict a peak in the
reflection and absorption at the frequency equals to the gap in the
quasiparticle spectrum. It is shown that the electron-hole pairing
results in an essential modification of the spectrum of surface TM
plasmons. We find that the optical TM mode splits into a low
frequency undamped branch and a high frequency damped branch. At
zero temperature the lower branch disappears. It is established that
the pairing does not influence the acoustic TM mode. It is also
shown that the pairing opens the frequency window in the subgap
range for the surface TE wave.
\end{abstract}


\maketitle
 \section{Introduction}

Electron-hole pairing in a system of two conducting layers with
oppositely charged carriers was predicted in  \cite{1,2}.  Such a
system below the Kosterlitz-Thouless transition temperature may
support a superflow of electron-hole pairs. It should result in zero
resistance  under a flow of oppositely directed and equal in modulus
electrical currents in the adjacent layers.

Zero counterflow resistance has not been observed yet, but a number of
experiments confirm that  the pairing occurs. The increase in the
interlayer drag resistance at low temperatures was detected in a
double quantum well in AlGaAs heterostructures \cite{3,4} and in
hybrid double layer systems comprising a single-layer (bilayer)
graphene in close proximity to a quantum well created in
GaAs\cite{5}. The upturn of the drag resistivity witnesses for the
approaching the paired state \cite{6,6a}. The electron-hole pairing
was also predicted\cite{7,8,9} for quantum Hall bilayers where the
overall filling factor of the lowest Landau level of two layers is
close to unity. The pairing was confirmed by transport
experiments\cite{10,11,12} where an exponential increase of the
longitudinal counterflow conductivity and the vanishing of Hall
resistance were discovered. Recent observation of a perfect
longitudinal interlayer drag in the Corbino disk geometry \cite{14}
also witnesses for the electron-hole pairing.

After the experimental discovery of graphene the double layer
graphene systems are considered as perspective ones for achieving
the electron-hole pairing and counterflow superconductivity
\cite{15,16,17}. Potentially, such systems have a number of
advantages. In graphene the electron and hole spectra coincide with
each other at low energies and the condition of nesting of the
electron and hole Fermi surfaces is fulfilled automatically. There
is no gap between the electron and hole subbands and the density of
carriers in the electron and hole layers can be easily controlled by
external gates. At last, conducting electrons in graphene do not
undergo a localization at low density of carriers.

The main obstacle in realizing the electron-hole pairing in graphene
 is the screening of the
Coulomb interaction  between electrons and holes\cite{18,18a}.
Strictly spearing, the same obstacle emerges for the carriers with
the quadratic spectrum, but in the latter case relatively high
temperature of pairing can be achieved in the low density limit. In
this limit electrons and  holes bind in small size pairs (smaller
than the average distance between the pairs) and the screening is
suppressed. The Dirac carriers are not coupled in small size pairs,
but they may form large size electron-hole pairs analogous to the
Cooper pairs. Then the screening is also suppressed. Depending on
the material parameters two situations are possible. If the bare
interaction is weak, the screening will be strong and the critical
temperature will be very low. Such a case was analyzed in
\cite{18a}. On the contrary, if the bare interaction is strong, the
pairing will be accompanied by an essential weakening of screening
and it will occur at high temperature \cite{19,20}. The latter
possibility can be realized if the Coulomb interaction strength
$\alpha_{eff}$ exceeds certain critical value $\alpha_c$ and the
distance between the graphene layers $d$ related to the inverse
Fermi wave number is small: $d k_F\lesssim 1$.
 The interaction strength (also
called the effective fine structure constant) is defined as
$\alpha_{eff}=e^2/\hbar \varepsilon v_F$, where $v_F$ is the Fermi
velocity and $\varepsilon$ is the dielectric constant of the
surrounding media.
 The parameter $\alpha_c$ grows up under increase in the number of Dirac components
$N$ (number of valleys times the number of spin components). For
graphene ($N=4$) the dynamical screening theory\cite{19} yields
$\alpha_c\simeq 1.5$. Thus  a double layer graphene system  in a
vacuum ($\varepsilon=1$ and $\alpha_{eff}=2.19$) can be in a paired
state at relatively high temperatures ($>1$ K). Other known
two-dimensional Dirac crystals (silicene, germanene \cite{21,22},
$\alpha$-graphyne\cite{agr}) have the same number of Dirac
components, but smaller $v_F$. Having the parameter $\alpha_{eff}$
approximately in two times larger than one for graphene,  double
layer silicene, germanene and $\alpha$-graphyne look more promising
for achieving high critical temperature. In particular, a double
layer silicene(germanene) system embedded into a  nanoporous or
nanostructured matrix with $\varepsilon \lesssim 2$\cite{np1,np} can
also demonstrate the electron-hole pairing at high temperature.
Another perspective system is a topological insulator (TI)\cite{24}.
Two opposite surfaces of a TI plate serve as two adjacent conducting
layers. The spectrum of the surface states of TI has an odd number
of Dirac cones.
 For $N=1$ the critical parameter
$\alpha_c\simeq 0.5$ \cite{19} and the pairing at high temperature
 is possible  in a TI plate embedded in a dielectric media
with $\varepsilon\lesssim 5$\cite{25}. One can also mention a
possibility \cite{pmf} of reducing the Fermi velocity in graphene
and shifting the system parameters into the strongly coupled pairing
regime by a periodic magnetic field.

Electron-hole pairing was also predicted for double layer graphene
systems subjected to a strong uniform magnetic field directed
perpendicular to the layers \cite{b1,k1,f1,f2}. The important
advantage of the quantum Hall state in  Dirac spectrum systems is a
large energy gap between the zeroth and the first  Landau level. The
screening in this case is considerable reduced even without pairing.
Double layer system made of graphene sheets with the band gap
(induced, for instance, by hydrogenation) is also a promising
candidate for obtaining the electron-hole superfluidity \cite{ke}.
The presence of the gap makes the formation of small size local
pairs possible. One can expect that in that case  the screening will
be of a minor importance.

A direct manifestation of  the electron-hole pairing would be an
observation of zero counterflow resistance. But since in
two-dimensional superfluid systems a flow causes  unbinding of
vortex pairs, a small voltage appears in any case. The presence of
areas where the pairing is suppressed (in the case of system
composed of puddles of the superconductive phase\cite{26}), can be
another source of nonzero resistance. Therefore, it is desirable to
have an independent indicator of pairing. For usual superconductors
the Meissner effect can serve as such an indicator. Diamagnetic
response of a double layer system with electron-hole pairing is very
small\cite{1,27} and it cannot be used for detection of the pairing.
The pairing may reveals itself in a considerable enhancement of
tunneling conductivity in the vicinity of the critical temperature
 \cite{ef-loz}. This phenomenon  interpreted as a fluctuational
internal Josephson effect is the general one for electron-hole
bilayers and it can be used as an indicator of pairing. The pairing
reduces screening that can be seen by measuring the electric field
of a test charge located near graphene layers. A strong increase of
this field under lowering in temperature can be considered as a
universal hallmark of the electron-hole pairing \cite{27}, but it
requires sensitive sensors with high spatial resolution.

In this paper   we consider microwave methods of indirect
observation of the pairing with reference to   a double layer
graphene system made of two gapless monolayer graphene sheets in
zero magnetic field. We study the influence of the electron-hole
pairing on the transmission and reflection characteristics of double
layer graphene systems. We also analyze how the pairing changes the
spectrum of surface plasmon-polaritons in double layer graphene
systems. In Sec. \ref{s2} we develop the approach based on the
generalized Nambu formalism. In difference with the original Nambu
approach\cite{28} we use the matrix Green's function of dimension
$4\times 4$. In Sec. \ref{s3} gauge invariant linear response
functions are obtained. In Sec. \ref{s4} an influence of the pairing
on the transmission, reflection and absorption of electromagnetic
waves in a double layer graphene system is analyzed. In Sec.
\ref{s5} surface TM and TE waves in a double layer graphene system
with the electron-hole pairing are studied.

\section{Extended Nambu formalism for a double layer graphene} \label{s2}

In the low-energy approximation the conducting electrons in a
graphene layer are described by the matrix Hamiltonian
\begin{equation}\label{1}
    H=\sum_{\bf{k},\alpha=\pm 1,\sigma=\pm \frac{1}{2}}\psi_{\mathbf{k}\alpha\sigma}^+
    \left[\hbar v_F\left( {k}_x \sigma_x + \alpha {k}_y
    \sigma_y\right)-\mu \sigma_0\right]\psi_{\mathbf{k}\alpha\sigma},
\end{equation}
where  $\sigma_i$ are the Pauli matrices, $\sigma_0$ is the
identity matrix, $\mathbf{k}$ is the wave vector counted from the
Dirac point, $\mu$ is the chemical potential, $\alpha$ is the valley
index, $\sigma$ is the spin index, $\psi_{\mathbf{k}\alpha\sigma}^+$
and $\psi_{\mathbf{k}\alpha\sigma}$ are the electron creation and
annihilation operators that have the spinor structure
\begin{equation}\label{1a}
   \psi_{\mathbf{k}\alpha\sigma}^+=\left(
                            c_A^+\quad
                            c_B^+
                                                  \right)_{\mathbf{k}\alpha\sigma},\quad
\psi_{\mathbf{k}\alpha\sigma}=\left(
                          \begin{array}{c}
                            c_A \\
                            c_B \\
                          \end{array}
                        \right)_{\mathbf{k}\alpha\sigma}.
\end{equation}
The components of the pseudospinors (\ref{1a}) are the operators of
creation and annihilation of electrons in the graphene sublattices A
and B.

A double layer system with  the electron-hole pairing is described
by the mean-field Hamiltonian presented in terms of 4-component
spinors \cite{17}
\begin{equation}\label{2}
    H_{mf}=\sum_{\bf{k},\alpha,\sigma}\Psi^+_{\bf{k}\alpha\sigma}\left[\hbar v_F\left( {k}_x
\hat{\gamma}_{0x} + \alpha{k}_y
    \hat{\gamma}_{0y}\right) -
    (\mu+\chi_{\mathbf{k}})\hat{\gamma}_{z0}
    -\Delta_{\mathbf{k}}\hat{\gamma}_{xz}\right]\Psi_{\bf{k}\alpha\sigma},
\end{equation}
where
\begin{equation}\label{3}
  \Psi_{\mathbf{k}\alpha\sigma}=\left(
           \begin{array}{c}
             c_{A1} \\
             c_{B1} \\
             c_{A2} \\
             c_{B2} \\
           \end{array}
         \right)_{\mathbf{k}\alpha\sigma}
\end{equation}
and $1$ and $2$ are the layer indexes. In the Hamiltonian (\ref{2})
$\hat{\gamma}_{\alpha\beta}$ are the $4\times 4$ matrices defined
through the direct product
\begin{equation}\label{2a}
    \hat{\gamma}_{\mu\nu}=\sigma_\mu \otimes \sigma_\nu.
\end{equation}
We imply the electron density in the layer 1 equals the hole density
in the layer 2. It corresponds to $\mu_1=-\mu_2=\mu$. Eq. (\ref{2})
contains the order parameter of  the electron-hole pairing
$\Delta_{\mathbf{k}}$ and the Hartree-Fock potential
$\chi_{\mathbf{k}}$. These quantities have to be found from the
self-consistence equations.

In the general case \cite{mink} the order parameter is a  $2\times
2$ matrix $\hat{\Delta}_{XX'}$, components of which describe the pairing
of an electron in the sublattice $X$ and  a hole in the  sublattice
$X'$. This matrix can be expressed through the Pauli matrices
$\hat{\Delta}=\sum_{\mu=0,x,y,z}\Delta_\mu\sigma_\mu$. In this paper
we consider the state, where only $\Delta_z$ is nonzero.
 It corresponds to the maximum energy gap in the
quasiparticle spectrum and the minimum of the ground state
energy\cite{17}.

We would note the difference between the formalism used here and the
approach \cite{15,16,19,20,27} in which the standard  Nambu
notation\cite{28} can be used\cite{27}. In the latter case the order
parameter is defined as an average
$\Delta_{\mathbf{k},\lambda}=\langle
c_{1,\mathbf{k},\lambda}^+c_{2,\mathbf{k},-\lambda}\rangle$, where
$c_{i,\mathbf{k},\lambda}^+$ is the operator of electron creation in
the state with the wave vector $\mathbf{k}$ in the Dirac subband
$\lambda$ in the layer $i$.  The single layer Hamiltonian (\ref{1})
written in terms of operators $c_{i,\mathbf{k},\lambda}$ has the
scalar form. Then the Hamiltonian of a double layer system can be
written in terms of two-component spinors.

In the extended Nambu formalism the Green's function is a $4\times
4$ matrix
\begin{equation}\label{4}
    \mathbf{G}(\mathbf{k},\omega,\alpha,\sigma)=\left[(\omega+i 0)\hat{\gamma}_{00}- v_F\left( {k}_x \hat{\gamma}_{0x}
 + \alpha {k}_y
    \hat{\gamma}_{0y}\right) + (\mu+{\chi}_\mathbf{k})\hat{\gamma}_{z0}
    +\Delta_{\mathbf{k}}\hat{\gamma}_{xz}\right]^{-1}.
\end{equation}
Here and below we set $\hbar=1$. The Green's function (\ref{4}) has
the valley and the spin indices. Each spin-valley component
contributes additively  to the response functions and these
contributions do not depend on $\alpha$ and $\sigma$. Therefore, one
can consider only one component and take into account the other
components by the factor $N=4$ in the final answer. Below we do the
computations for $\alpha=1$ component and omit the spin and valley
indexes.

It is convenient to present the Green's function (\ref{4}) in the
form
\begin{equation}\label{4n}
    \mathbf{G}(\mathbf{k},i \omega)=\sum_{\lambda=\pm1}
\left(\frac{\mathbf{A}_{\lambda,\mathbf{k}}+\mathbf{B}_{\lambda,\mathbf{k}}}{i\omega-E_{\lambda,\mathbf{k}}}
+\frac{\mathbf{A}_{\lambda,\mathbf{k}}-\mathbf{B}_{\lambda,\mathbf{k}}}{i\omega+E_{\lambda,\mathbf{k}}}\right),
\end{equation}
where
\begin{equation}\label{4a}
    E_{\lambda,\mathbf{k}}=
   \sqrt{\xi_{\lambda,\mathbf{k}}^2+\Delta_{\mathbf{k}}^2},
\end{equation}
and $\xi_{\lambda,\mathbf{k}}=\lambda v_F k -\mu-
\chi_{\mathbf{k}}$. The matrices $\mathbf{A}_{\lambda,\mathbf{k}}$,
$\mathbf{B}_{\lambda,\mathbf{k}}$ are expressed through the
$\hat{\gamma}_{\mu\nu}$ matrices:
\begin{equation}\label{5n}
\mathbf{A}_{\lambda,\mathbf{k}}=\frac{1}{4}\left(\hat{\gamma}_{00}
+\lambda\hat{\gamma}_{zx}\cos\theta_{\mathbf{k}}
+\lambda\hat{\gamma}_{zy}\sin\theta_{\mathbf{k}}\right),
\end{equation}
\begin{equation}\label{6n}
\mathbf{B}_{\lambda,\mathbf{k}}=\frac{\Delta_{\mathbf{k}}}{4
E_{\lambda,\mathbf{k}}}\left(\hat{\gamma}_{xz}
+\lambda\hat{\gamma}_{yy}\cos\theta_{\mathbf{k}}
-\lambda\hat{\gamma}_{yx}\sin\theta_{\mathbf{k}}\right)+
\frac{\xi_{\lambda,\mathbf{k}}}{4
E_{\lambda,\mathbf{k}}}\left(\hat{\gamma}_{z0}
+\lambda\hat{\gamma}_{0x}\cos\theta_{\mathbf{k}}
+\lambda\hat{\gamma}_{0y}\sin\theta_{\mathbf{k}}\right),
\end{equation}
where  $\theta_\mathbf{k}$ is the angle between the wave vector
$\mathbf{k}$ and the $x$ axes.

The interaction part of the Hamiltonian written in terms of
four-component spinors (\ref{3}) reads
\begin{eqnarray}\label{d1}
H_{int}=\frac{1}{4 S}\sum_{\mathbf{k},\mathbf{k}',\mathbf{q}}
\Bigg[V_+(q):\Psi^+_{\bf{k}+\mathbf{q}}\hat{\gamma}_{00}\Psi_{\bf{k}}\Psi^+_{\bf{k}'-\mathbf{q}}
\hat{\gamma}_{00} \Psi_{\bf{k}'}: +V_-(q):\Psi^+_{\bf{k}+\mathbf{q}}
\hat{\gamma}_{z0}
\Psi_{\bf{k}}\Psi^+_{\bf{k}'-\mathbf{q}}\hat{\gamma}_{z0}\Psi_{\bf{k}'}:\Bigg]\cr
+\sum_{\bf{k}}\Psi^+_{\bf{k}}\left(\chi_{\mathbf{k}}\hat{\gamma}_{z0}
+\Delta_{\mathbf{k}}\hat{\gamma}_{xz}\right)\Psi_{\bf{k}},
\end{eqnarray}
where $S$ is the area of the system, the notation
$:\Psi^+\hat{\gamma}\Psi\Psi^+\hat{\gamma}\Psi:$ means the normal
ordering, $V_{\pm}(q)=V_S(q)\pm V_D(q)$, and $V_S(q)$, $V_D(q)$ are the Fourier-components of the intralayer and interlayer Coulomb
interaction, respectively. We specify the case of a uniform
dielectric environment that corresponds to the same intralayer
interaction potential $V_S$ in both layers.  The second sum in Eq.
(\ref{d1}) compensates the mean-field terms in the Hamiltonian
(\ref{2}).

The self-consistence condition requires the nullifying of the lowest
order self-energy correction to the mean-field Green's function
(\ref{4}). The self-energy
 $\bm{\Sigma}(\mathbf{k},\omega)$ is the $4\times 4$ matrix and the
 self-consistence equation has the matrix form equivalent to 16
 scalar equations. Most of them are satisfied by symmetry.
 Nontrivial ones are the self-consistence equation for the
 Hartree-Fock potential $\chi_\mathbf{k}$ and the equation for the order parameter $\Delta_{\mathbf{k}}$.
The main effect of the Hartree-Fock potential is an unessential shift of
the chemical potential that can be included in the definition of
$\mu$. The order parameter satisfies the equation
\begin{equation}\label{4b}
\Delta_{\mathbf{k}}=\frac{1}{S}\sum_{\mathbf{k}'}\sum_{\lambda}
\frac{V_{D}(|\mathbf{k}-\mathbf{k}'|)}{2}
\frac{\Delta_{\mathbf{k}'}}{2
E_{\lambda,\mathbf{k}'}}\tanh\frac{E_{\lambda,\mathbf{k}'}}{2 T},
\end{equation}
where $T$ is the temperature. We note that Eq. (\ref{4b}) does not
contain the factor with cosine in difference with the
self-consistence equation used in \cite{15,16,19,20,27}. The origin
of this difference is that the order parameter $\Delta_{\mathbf{k}}$
introduced in the Hamiltonian (\ref{2}) does not depend on
$\lambda$. In the approach \cite{15,16,19,20,27} the order parameter
depends on $\lambda$. In that case the self-consistence equation has
the form
\begin{equation}\label{4b1}
  \Delta_{\lambda,\mathbf{k}}=\frac{1}{S}\sum_{\mathbf{k}'}\sum_{\lambda'}
\frac{V_{D}(|\mathbf{k}-\mathbf{k}'|)}{2}
[1+\lambda\lambda'\cos(\theta_{\mathbf{k}'}-\theta_\mathbf{k})]
\frac{\Delta_{\lambda',\mathbf{k}'}}{2
E_{\lambda',\mathbf{k}'}}\tanh\frac{E_{\lambda',\mathbf{k}'}}{2 T}.
\end{equation}
The consistence of Eq. (\ref{4b1})  with the condition
$\Delta_{+1,\mathbf{k}}=\Delta_{-1,\mathbf{k}}$ requires zero
contribution of the term with cosine into the integral in Eq.
(\ref{4b1}). Then Eq. (\ref{4b1}) reduces to Eq. (\ref{4b}).

The screening can be taken into account by replacing the bare interaction $V_D$
with the screened one. In the random phase approximation the
Fourier-component of the screened interaction reads
\begin{equation}\label{d2}
    V_D^{scr}(\mathbf{k},\omega)=\frac{1}{2}\left(\frac{V_+({k})}{1-V_+({k})
{\Pi}_{+,00}(\mathbf{k},\omega)}-
\frac{V_-({k})}{1-V_-({k}){\Pi}_{-,00}(\mathbf{k},\omega)}\right),
\end{equation}
where $\Pi_{\pm,00}(\mathbf{k},\omega)$ are the density-density
response functions defined below.

In the static screening approximation the quantity
$V_{D}(\mathbf{k})$ in Eq. (\ref{4b}) is replaced with
$V_D^{scr}(\mathbf{k},0)$. In the dynamical screening approximation
  the self-consistence equation is modified to \cite{19}
\begin{equation}\label{4h}
\Delta_{\mathbf{k}}(i\Omega)=\frac{T}{S}  \sum_{{n=-\infty}}^\infty
\sum_{\mathbf{k}'}\sum_{\lambda}
\frac{V^{scr}_{D}[\mathbf{k}-\mathbf{k}',i(\Omega-\omega_{n})]}{2}
\frac{\Delta_{\mathbf{k}'}(i\omega_{n})}{E_{\lambda,\mathbf{k}'}^2+\omega_{n}^2},
\end{equation}
where $\omega_n=\pi T (2n+1)$ are the odd Matsubara frequencies.

\section{Ward identity and gauge invariance of the response functions}
\label{s3}

The interaction of the double layer graphene system with the
electromagnetic field is described by the Hamiltonian
\begin{equation}\label{5}
    H_A=-\frac{1}{2S c}\sum_{\mu=0,x,y}\sum_\mathbf{q}\left[\hat{j}_{+,\mu}(\mathbf{q})
    A_{+,\mu}(-\mathbf{q},t)+
    \hat{j}_{-,\mu}(\mathbf{q})A_{-,\mu}(-\mathbf{q},t)\right],
\end{equation}
where $\hat{j}_{\pm,0}$ are the charge density operators,
$\hat{j}_{\pm,i}$ - are the current density operators ($i=x,y$),
$A_{\pm,i}$ - are the vector potentials, $A_{\pm,0}=-c\varphi_{\pm}$,
$\varphi_{\pm}$ are the scalar potentials, and $c$ is the light
velocity. The index $"+"$ ($"-"$) notates  the sum (difference) of
the corresponding quantities in the layers 1 and 2. The charge and
current density operators are given by the equation
\begin{equation}\label{6}
    \hat{j}_{\pm,\mu} (\mathbf{q})
    =e\sum_{\mathbf{k}}\Psi^+_{\mathbf{k}+\mathbf{q}}
\bm{\gamma}^{\pm}_\mu\Psi_{\mathbf{k}},
\end{equation}
where the vertexes have the matrix form
\begin{equation}\label{6a}
    \bm{\gamma}^+_0=\hat{\gamma}_{00}, \quad \bm{\gamma}^-_0=\hat{\gamma}_{z0},
\quad \bm{\gamma}^+_i= v_F\hat{\gamma}_{0i}, \quad \bm{\gamma}^-_i=
v_F\hat{\gamma}_{zi}.
\end{equation}

In the linear response approximation the currents $j_{\pm,\mu}$ are the
linear functions of the potentials $A_{\pm,\nu}$:
\begin{equation}\label{7a}
     j_{\pm,\mu}
(\mathbf{q},\omega)=-\frac{e^2}{c}\Pi_{\pm,\mu\nu}(\mathbf{q},\omega)
A_{\pm,\nu}(\mathbf{q},\omega).
\end{equation}
The response functions $\Pi_{\pm,\mu\nu}(\mathbf{q},\omega)$ are
obtained as the analytical continuation of the imaginary frequency
current-current correlators
\begin{equation}\label{7b}
    \Pi_{\pm,\mu\nu}(\mathbf{q},i\omega)=-\frac{N}{2 S
    e^2}\int_0^\beta d\tau
e^{i\omega\tau}\langle T_\tau
\hat{j}_{\pm,\mu}(\mathbf{q},\tau)\hat{j}_{\pm,\nu}(-\mathbf{q},0)\rangle,
\end{equation}
where $N=4$ is the number of Dirac components,
$\hat{j}_{\pm,\mu}(\mathbf{q},\tau)=e^{H_{mf}\tau}\hat{j}_{\pm,\mu}(\mathbf{q})e^{-H_{mf}\tau}$
, $\beta=1/T$, and $T_\tau$ is the $\tau$ ordering operator.

Neglecting the interaction (\ref{d1}) one obtains the mean-field
response functions
\begin{eqnarray}\label{7c}
    \Pi_{\pm,\mu\nu}(\mathbf{q},i\Omega)=
    \frac{2T}{S}\sum_{n=-\infty}^{\infty}
\sum_{\mathbf{k}} \mathrm{Tr}[\bm{\gamma}^\pm_\mu
\mathbf{G}(\mathbf{k}+\mathbf{q},i\omega_n+i\Omega)\bm{\gamma}^\pm_\nu
\mathbf{G}(\mathbf{k},i\omega_n)].
\end{eqnarray}

It is well known \cite{28,29}  that the mean-field response
functions in the Bardeen-Cooper-Schrieffer (BCS) theory  are not
gauge invariant. It is the result of that in the mean-field
approximation the interaction is not taken into account properly.
The interaction is included in the self-energy part while the
renormalization of vertexes is neglected.

Any gauge transformation of the potentials $A_{\pm,\mu}$ should
leave the currents ({\ref{7a}) unchanged. This condition corresponds
to the following equation for the response functions
\begin{equation}\label{10}
   i\omega\Pi_{\pm,\mu 0}(\mathbf{q},i\omega)-
\sum_{i=x,y}q_i \Pi_{\pm,\mu i}(\mathbf{q},i\omega)=0.
\end{equation}
If  $\Pi_{\pm,\mu \nu}$ are not  gauge invariant,  the current
(\ref{7a}) will not satisfy the continuity equation. Then
considering an electromagnetic problem we would obtain one answer
with the use of the boundary condition for the electric displacement
field and another answer with the use of the boundary condition for
the magnetic field. It would make the theory inconsistent.
Therefore, for the problems
 we study in the next two sections the gauge invariance of the response
functions is mandatory.

Gauge invariance can be restored by dressing the vertexes. The
response functions
\begin{equation}\label{7d}
    \Pi_{\pm,\mu\nu}(\mathbf{q},i\Omega)=
    \frac{2 }{S}T\sum_{n=-\infty}^{\infty}
\sum_{\mathbf{k}} \mathrm{Tr}[\bm{\gamma}^\pm_\mu
\mathbf{G}(\mathbf{k}+\mathbf{q},i\omega_n+i\Omega)\bm{\Gamma}^\pm_\nu
(\mathbf{k}+\mathbf{q},\mathbf{k},i\omega_n+i\Omega,i\omega_n)
\mathbf{G}(\mathbf{k},i\omega_n)]
\end{equation}
will be gauge invariant if the dressed vertexes
$\bm{\Gamma}^\pm_{i}$ satisfy the generalized Ward identity
\begin{eqnarray}\label{8a}
   \sum_{i=x,y} q_i \bm{\Gamma}^\pm_{i}(\mathbf{k}+\mathbf{q},\mathbf{k},i\omega_n+i\Omega,i\omega_n)
    - i\Omega
   \bm{\Gamma}^\pm_{0}(\mathbf{k}+\mathbf{q},\mathbf{k},i\omega_n+i\Omega,i\omega_n)
=\bm{\gamma}^\pm_{0}
\bm{G}^{-1}(\mathbf{k},i\omega_n)-\bm{G}^{-1}(\mathbf{k}+\mathbf{q},i\omega_n+i\Omega)\bm{\gamma}^\pm_{
0}.
\end{eqnarray}
This statement can be proven by the direct substitution of Eqs.
(\ref{7d}) and (\ref{8a}) into Eq. (\ref{10}). One can also prove
that, as in the BCS theory \cite{29}, the ladder approximation
yields the vertexes that satisfy the Ward identity. For finding the
vertex function in the ladder approximation one should solve a
matrix integral equation. It is extremely cumbersome problem.
Fortunately, the vertexes that satisfy the identity (\ref{8a}) can
be found in a simpler way.

In the absence of interaction the Hartree-Fock potential
$\chi_{\mathbf{k}}$ and the order parameter $\Delta_{\mathbf{k}}$
are equal to zero and the Ward identity is fulfilled for the bare
vertexes $\bm{\Gamma}^\pm_\mu=\bm{\gamma}^\pm_\mu$.  If the
Hartree-Fock potential is taken into account only as a shift of the
chemical potential, the bare vertexes will  satisfy the Ward
identity in the normal state ($\Delta_{\mathbf{k}}=0$).

For the paired state we consider the gauge invariance problem in the
constant gap ($\Delta_{\mathbf{k}}=\Delta$) and constant
Hartree-Fock potential  approximation. The constant Hartree-Fock
potential is included below into the definition of $\mu$. Then Eq.
(\ref{8a}) reduces to
\begin{eqnarray}\label{18a}
    q_x \bm{\Gamma}^+_{x}+q_y \bm{\Gamma}^+_{y}  - i\Omega
   \bm{\Gamma}^+_{0}
=v_F(\hat{\gamma}_{0x} q_x+\hat{\gamma}_{0y} q_y)-i \Omega
\hat{\gamma}_{00},
\end{eqnarray}
\begin{eqnarray}\label{18b}
    q_x \bm{\Gamma}^-_{x}+q_y \bm{\Gamma}^-_{y}  - i\Omega
   \bm{\Gamma}^-_{0}
=v_F(\hat{\gamma}_{zx} q_x+\hat{\gamma}_{zy} q_y)-i \Omega
\hat{\gamma}_{z0}+2i \hat{\gamma}_{yz}\Delta.
\end{eqnarray}

One can see that the bare vertexes
$\bm{\Gamma}^+_{\mu}=\bm{\gamma}^+_{\mu}$ satisfy the Ward identity
(\ref{18a}). Thus  in the constant gap approximation the mean-field
response functions $\Pi_{+,\mu\nu}$ (\ref{7c}) are gauge invariant.
The bare vertexes $\bm{\Gamma}^-_{\mu}=\bm{\gamma}^-_{\mu}$ do not
satisfy the Ward identity. The  vertexes $\bm{\Gamma}^-_{\mu}$ that
satisfy Eq. (\ref{18b}) can be found as follows.  One can see from
Eq. (\ref{18b}) that the vertex function $\bm{\Gamma}^-_{\mu}$
depends only on $\mathbf{q}$ and $\Omega$.  The solution of Eq.
(\ref{18b}) is presented in the form
$\bm{\Gamma}_\mu^-=\bm{\gamma}^-_{\mu}+\tilde{\bm{\Gamma}}_\mu^-(\mathbf{q},i\Omega)$,
where $\tilde{\bm{\Gamma}}_\mu^-(\mathbf{q},i\Omega)$ satisfies the
equation
\begin{eqnarray}\label{18c}
    q_x \tilde{\bm{\Gamma}}^-_{x}+q_y \tilde{\bm{\Gamma}}^-_{y}  - i\Omega
   \tilde{\bm{\Gamma}}^-_{0}
=2i \hat{\gamma}_{yz}\Delta.
\end{eqnarray}
 The functions
$\tilde{\bm{\Gamma}}_\mu^-(\mathbf{q},i\Omega)$ should have a pole
at $\Omega\to 0$ and $q\to 0$. This pole corresponds to the
Anderson-Bogoliubov(AB) mode \cite{30,31}. The AB mode is connected
with fluctuations of the phase of the order parameter. In a
three-dimensional superconductor the spectrum of this mode is
$\Omega_q=v_F q/\sqrt{3}$. A genuine AB mode emerges only in neutral
superfluids. In charged system the AB mode is coupled to the scalar
potential and it renormalizes the electromagnetic response function.
In the bilayer system the phase of the order parameter is coupled to
the  potential $\varphi_-$ and the AB mode renormalizes the response
functions $ \Pi_{-,\mu\nu}$. One can show that in a two dimensional
system the spectrum of the AB mode is modified as $\Omega_q=v_F
q/\sqrt{2}$. Basing on the arguments given above, we seek for a
solution of Eq. (\ref{18c}) in the form
\begin{equation}\label{19}
  \tilde{\bm{\Gamma}}_\mu^-=-i\hat{\gamma}_{yz}\Delta
\frac{f_\mu(\mathbf{q},i\Omega)}{(i\Omega)^2-s^2 q^2},
\end{equation}
where $s=v_F/\sqrt{2}$. The functions $f_\mu(\mathbf{q},i\Omega)$
are assumed to be regular at $\Omega\to 0$ and $q\to 0$. Then, in
the linear in $q$ and $\Omega$ order we find $f_{i}=2 s^2 q_i$ and
$f_{0}=2 i \Omega$. Finally, we obtain the following renormalized
vertexes
\begin{eqnarray} \label{18f}
   \bm{\Gamma}_0^-= \bm{\gamma}^-_{0} -2i\hat{\gamma}_{yz}\Delta \frac{i\Omega}{(i\Omega)^2-s^2
q^2},\cr
    \bm{\Gamma}_i^-= \bm{\gamma}^-_{i}-2i\hat{\gamma}_{yz}\Delta s^2 \frac{q_i}{(i\Omega)^2-s^2
q^2}.
\end{eqnarray}
The  vertexes (\ref{18f}) satisfy the Ward identity and  guarantee
the gauge invariance of the response functions (\ref{7d}). We
emphasize that
 Eqs. (\ref{18f}) are derived in the low frequency
long wavelength limit.

\section{Reflection, transmission and absorption in the terahertz range}
\label{s4}

 The coefficient of transmission $T$ for an electromagnetic
wave going through an undoped graphene  is almost frequency
independent in a wide  frequency range. The absorption is caused by
the transition between filled  and empty states in the Dirac cones.
The coefficient of absorption is equal to $A=\pi\alpha$, where
$\alpha=e^2/\hbar c$ is the fine structure constant\cite{l1,l2,l3}.
The reflectivity of graphene $R$ is extremely small, proportional to
$\alpha^2$. In the electron
 doped graphene at zero temperature the low energy
states (counted from the Dirac point)  are all filled. Therefore,
 there is no absorption \cite{l4} in the frequency range $1/\tau\ll \omega < 2 \mu$, where
$\tau$ is the relaxation time.
 A double layer electron-hole graphene system with spatially
 separated layers should demonstrate the same behavior in the normal state. If the
 tunneling between the layers is negligible, an electron
from one layer cannot transit to the empty state in the other layer
and the layers contribute independently. In the  paired state the
quasiparticles do not belong to a given layer. At the same time, the
quasiparticle spectrum has a gap $2\Delta$. Therefore, one can
expect the shifting of the absorption edge from $\omega=2\mu$ to
$\omega=2 \Delta$.

To obtain the frequency dependence of the transmission, reflection,
and absorption coefficients we consider a $p$-polarized incident
wave (the magnetic field of the wave is
 parallel to the graphene layers).  At the frequency of the wave $\omega\sim \mu$
and for the interlayer distance $d < k_F^{-1}$ the ratio of the
interlayer distance to the wavelength is $d/\lambda\lesssim
10^{-3}$. Since
 $d$ is much smaller than the wavelength one can consider a double
layer graphene system   as a single boundary. The electric current
at that boundary
 is the sum of currents in two graphene layers
$\mathbf{j}_{+}$. The parallel current conductivity tensor
$\hat{\sigma}_+$ relates the current $\mathbf{j}_+$ with the
electric field $\mathbf{E}_+$: $j_{+,x}=\sigma_{+,xx}E_{+,x}$. For
$d/\lambda\ll 1$ the fields $E_{1,x}=E_{2,x}={E}_x$, where the $E_x$
is the tangential component of the electric field at the boundary,
and $E_{+,x}=2 {E}_x$.

Let the incident wave has the wave vector $\mathbf{k}=(k_x,0,k_z)$.
Then the boundary condition for the magnetic field is
\begin{equation}\label{bc}
   H_{1,y}-H_{2,y}=-\frac{4\pi}{c}j_{+,x}=-\frac{8\pi}{c}\sigma_{+,xx}
   E_x.
\end{equation}
Here the boundary at  $z=0$ is implied, and the indexes 1 and 2
stand for the $z>0$ and $z<0$ half-spaces.

 The boundary condition (\ref{bc}) together with Maxwell equations determine the following
relations between the amplitudes of the electric  field of the
incident ($i$), reflected ($r$) and transmitted ($t$) waves
\begin{eqnarray} \label{301}
  E_r = E_i \left|\frac{\frac{4\pi \sigma_{+,xx}(\mathbf{k}_{pl},\omega)\cos \theta}{c}}
  {1+ \frac{4\pi \sigma_{+,xx}(\mathbf{k}_{pl},\omega)\cos
  \theta}{c}}\right|,\quad
  E_t &=& E_i \left|\frac{1}{1+ \frac{4\pi \sigma_{+,xx}(\mathbf{k}_{pl},\omega)
  \cos \theta}{c}}\right|,
\end{eqnarray}
where $\theta$ is the incident angle. The amplitudes of the magnetic
components of the waves satisfy the same relations.  In Eqs.
(\ref{301}) $\mathbf{k}_{pl}=\mathbf{i}_x k_x $ is the tangential
component of the wave vector, and $\mathbf{i}_x$ is the unit vector
along the $x$ axis. From the relations (\ref{301}) one finds the
transmission, reflection, and absorption coefficients for the normal
incidence ($\theta=0$)
\begin{eqnarray} \label{20a}
  {T} = \left|\frac{1}
  {1+\frac{4\pi \sigma_{+}(\omega)}{c}}\right|^2,\quad
  {R} = \left|\frac{\frac{4\pi \sigma_+(\omega)}{c}}
  {1+\frac{4\pi \sigma_{+}(\omega)}{c}}\right|^2,
   \quad {A}=1-{R}
-{T},
\end{eqnarray}
where $\sigma_{+}(\omega)=\sigma_{+,xx}(0,\omega)$ is the uniform
high-frequency conductivity. Using the boundary condition for the
normal component of the electric field ($E_{1,z}-E_{2,z}=4\pi
\rho_{+}$) one obtains the same expressions for  $T$, $R$ and $A$.
To get them one should take into account the continuity equation and
the relation
\begin{equation}\label{21}
\Pi_{+,xx}(q \mathbf{i}_x,\omega)=\frac{\omega^2}{q^2}\Pi_{+,00}(q
\mathbf{i}_x ,\omega)
\end{equation}
that comes from Eq. (\ref{10}).

The conductivity $\sigma_{+}(\omega)$  is given by the response
function $\Pi_{+,xx}$
\begin{equation}\label{20}
    \sigma_+(\omega)=ie^2\frac{\Pi_{+,xx}(0,\omega)}{\omega}.
\end{equation}
The relation (\ref{21}) allows to get $\sigma_{+}(\omega)$ from the
response function $\Pi_{+,00}$, as well:
\begin{equation}\label{22}
\sigma_{+}(\omega)=i{\omega}e^2\lim_{q\to 0}
\frac{\Pi_{+,00}(\mathbf{q} ,\omega)}{q^2}.
\end{equation}

 Using the
expansion (\ref{4n}), computing the traces, summing over the
imaginary frequencies, and doing the analytical continuation we get
from (\ref{7c}) the following expressions for response functions
\begin{eqnarray}\label{22a}
\Pi_{+,00}(\mathbf{q},\omega)=- 4 \sum_{\lambda,\lambda'}\int
\frac{d \mathbf{k}}{(2\pi)^2}
F^{00}_{\lambda,\lambda',\mathbf{k},\mathbf{q}}\Big[
P^{00}_{\lambda,\lambda',\mathbf{k},\mathbf{q}}\frac{(1-f'
-f)(E+E')}
{(E+E')^2-\omega^2}+L^{00}_{\lambda,\lambda',\mathbf{k},\mathbf{q}}\frac{(f'-f
)(E-E')} {(E-E')^2-\omega^2}\Big],
\end{eqnarray}
\begin{eqnarray}\label{23}
\Pi_{+,xx}(\mathbf{q},\omega)=-4  v_F^2\sum_{\lambda,\lambda'}\int
\frac{d \mathbf{k}}{(2\pi)^2}
F^{xx}_{\lambda,\lambda',\mathbf{k},\mathbf{q}}\Big[
P^{xx}_{\lambda,\lambda',\mathbf{k},\mathbf{q}}\frac{(1-f'
-f)(E+E')}
{(E+E')^2-\omega^2}+L^{xx}_{\lambda,\lambda',\mathbf{k},\mathbf{q}}\frac{(f'-f
)(E-E')} {(E-E')^2-\omega^2}\Big].
\end{eqnarray}
 Here
and below the abbreviated notations $E\equiv
E_{\lambda,\mathbf{k}}$, $E'\equiv
E_{\lambda',\mathbf{k}+\mathbf{q}}$,
$f\equiv(\exp(E_{\lambda,\mathbf{k}}/T)+1)^{-1}$, and
$f'\equiv(\exp(E_{\lambda',\mathbf{k}+\mathbf{q}}/T)+1)^{-1}$ are
used.  The factors $F$, $P$ and $L$ in  Eqs. (\ref{22a}) and
(\ref{23}) have the form
\begin{equation}\label{24}
   F^{00}_{\lambda,\lambda',\mathbf{k},\mathbf{q}}=
   \frac{1+\lambda \lambda'\cos(\theta_{\mathbf{k}+\mathbf{q}}-
   \theta_\mathbf{k})
   }{2}, \quad F^{xx}_{\lambda,\lambda',\mathbf{k},\mathbf{q}}=
   \frac{1+\lambda \lambda'\cos(\theta_{\mathbf{k}+\mathbf{q}}+
   \theta_\mathbf{k})
   }{2},
   \end{equation}
\begin{eqnarray} \label{25}
  P^{00}_{\lambda,\lambda',\mathbf{k},\mathbf{q}}
  =
  \frac{1}{2}\left(1-\frac{\xi_{\lambda,\mathbf{k}}\xi_{\lambda',\mathbf{k}+\mathbf{q}}+\Delta^2}
  {E_{\lambda,\mathbf{k}}E_{\lambda',\mathbf{k}+\mathbf{q}}}
  \right), \quad
  L^{00}_{\lambda,\lambda',\mathbf{k},\mathbf{q}} = \frac{1}{2}
  \left(1+\frac{\xi_{\lambda,\mathbf{k}}\xi_{\lambda',\mathbf{k}+\mathbf{q}}+\Delta^2}
  {E_{\lambda,\mathbf{k}}E_{\lambda',\mathbf{k}+\mathbf{q}}}
\right),
  \cr
  P^{xx}_{\lambda,\lambda',\mathbf{k},\mathbf{q}} = \frac{1}{2}
  \left(1-\frac{\xi_{\lambda,\mathbf{k}}\xi_{\lambda',\mathbf{k}+\mathbf{q}}-\Delta^2}
  {E_{\lambda,\mathbf{k}}E_{\lambda',\mathbf{k}+\mathbf{q}}} \right), \quad
  L^{xx}_{\lambda,\lambda',\mathbf{k},\mathbf{q}} =\frac{1}{2}
  \left(1+\frac{\xi_{\lambda,\mathbf{k}}\xi_{\lambda',\mathbf{k}+\mathbf{q}}-\Delta^2}
  {E_{\lambda,\mathbf{k}}E_{\lambda',\mathbf{k}+\mathbf{q}}}
\right).
\end{eqnarray}
 The factors (\ref{25}) coincide with the coherence  factors
in the BCS theory \cite{28,29}.

We note that the four-component spinor formalism yields the same
result for the response functions $\Pi_{+,\mu\nu}$ as the
two-component spinor formalism\cite{27}.  But the important
advantage of the formalism used in this paper is that it is
compatible with the algorithms of checking and restoring of the
gauge invariance developed in the BCS theory.

The integral in the expression for $\Pi_{+,xx}(\mathbf{q},\omega)$
diverges. This divergence is unphysical one. It is connected with
the linear approximation for the spectrum at large $k$. The same
problem emerges for the monolayer graphene\cite{34,35}.
 To escape this problem one should regularize the
expression for $\Pi_{+,xx}(\mathbf{q},\omega)$.  Since the constant
magnetic field cannot induce electrical currents in the normal
system,  the regularized response function $\Pi_{+,xx}^r$ should be
zero at $\omega=0$ and $\Delta=0$ . This condition is fulfilled if
one adds to the function (\ref{23}) the compensating term that
depends only on the cutoff wave vector $k_m$:
\begin{equation}\label{ne100}
  \Pi_{+,xx}^r=\Pi_{+,xx} + \frac{v_F k_m}{\pi }.
\end{equation}
Here $\Pi_{+,xx}$ is given by Eq. (\ref{23}), where the integral
over $k$ is taken with the upper cutoff  $k_m$.

The rule of integration of singularities
 in Eqs. (\ref{22a}), (\ref{23}) is fixed by
   the standard substitution $\omega \to \omega+i\eta$, where $\eta=+0$.
 For the computation we imply a finite
 $\eta=10^{-3}\mu$ considering it as  a phenomenological
scattering parameter.

We have computed $\sigma_+(\omega)$ from the regularized response
function $\Pi^r_{+,xx}(q \mathbf{i}_x,\omega)$  and from the
response function $\Pi_{+,00}(q \mathbf{i}_x,\omega)$  and have got
the same result. It confirms that the response functions obtained
numerically are gauge invariant.

The real and imaginary parts of the conductivity $\sigma_+(\omega)$
at the temperature $T=0.1\mu$ and two different $\Delta$ are
presented in Fig. \ref{fig1}. For the comparison the conductivity in
the normal state is also shown. The frequency dependence of the
transmission, reflection and absorption coefficients for the system
with pairing and for the normal system are given in Fig. \ref{fig2}.
One can see that the pairing results in the appearance of a sharp
peak in the reflection and absorption at $\omega=2 \Delta$. This
peak is accompanied with a deep minimum in the transmission. Below
the gap ($\omega<2\Delta$) the system with pairing is completely
transparent, while at $\omega>2\Delta$ it demonstrates significant
absorption.  In this range of frequencies the absorption coefficient
decreases under increase in frequency up to
$\omega=2\sqrt{\Delta^2+\mu^2}$. At that frequency a step-like
increase of the absorption coefficient  occurs. It is connected with
opening of an additional channel of absorption caused by the
transition between $\lambda=+1$ and $\lambda=-1$ states (\ref{4a}).
Normal systems demonstrate  similar step-like features at
$\omega=2\mu$, but at much lower temperatures. At $T=0.1\mu$ a
step-like peculiarity is completely smeared out in the normal system
(Fig. \ref{fig2}, left panel). In Fig. \ref{fig2} the spectral
characteristics are shown for $\Delta=0.5\mu$ and $\Delta=0.2\mu$.
One can see that lowering of $\Delta$ results  in a shift of the
peak position. It the same time, the height and the width of the
peaks remains practically unchanged. The computations show that the
peaks caused by pairing are detectable already at $\Delta=0.01\mu$.

The gap in the excitation spectrum may emerge not only due to
electron-hole pairing, but due to the interlayer tunneling, as well.
Absorption and reflection coefficients are not sensitive to the gap
origin and the tunneling  can mimic the effect of pairing. To
identify the origin of the gap one can take into account that the
gap caused by pairing depends on temperature. For the systems with
pairing we expect a red shift of the peak and their disappearance
under increase in temperature.

Strong concentration mismatch of graphene layers may also modify the
spectral characteristics of double layer systems. Such a mismatch
results in an appearance of two step-like singularity at
$\omega=2\mu_1$ and $\omega=2\mu_2$. The amplitude and the shape of
these singularities differ significantly from  ones for the peaks
caused by pairing. Besides, in the normal system the step-like
singularities are smeared out at rather small temperatures. It
allows to distinguish the effect of the mismatch and of the pairing.

\begin{figure}
\begin{center}
\includegraphics[width=6cm]{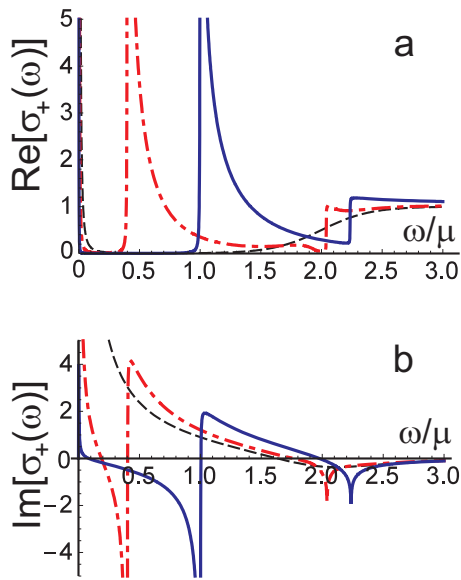}
\end{center}
\caption{Frequency dependence of the parallel current conductivity,
the real (a) and imaginary (b) parts, in the paired state with
$\Delta=0.5\mu$ (solid line) and $\Delta=0.2\mu$ (dash-dotted line),
and in the normal state (dashed line). The conductivity is given in
$e^2/4\hbar$ units. The temperature $T=0.1 \mu$.} \label{fig1}
\end{figure}

\begin{figure}
\begin{center}
\includegraphics[width=10cm]{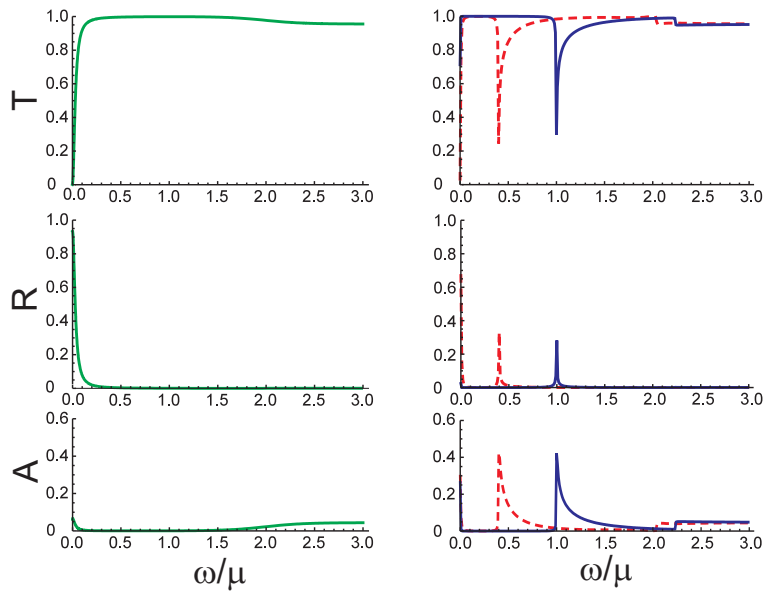}
\end{center}
\caption{Frequency dependence of the transmission ($T$), reflection
($R$) and absorption ($A$) coefficients for the normal incidence.
Left panel, the double layer graphene system in the normal state;
right panel, the same system in the paired state with $\Delta=0.5
\mu$ (solid curves) and $\Delta=0.2 \mu$ (dashed curves). The
temperature $T=0.1\mu$. } \label{fig2}
\end{figure}

Concluding the section we note that a similar problem, an impact of
an excitonic gap in the high-frequency conductivity, was considered
in \cite{l5} with reference to a single graphene layer where the
interaction may cause an opening of a gap in the quasiparticle
spectrum in zero or in a finite magnetic field \cite{l6}. It was
shown in \cite{l5} that the low $\omega$ part of the interband
contribution to the conductivity cuts off at $2\mu$ or $2\Delta$
whichever is the largest. For the bilayer system with the pairing we
predict different behavior. The cutoff frequency is the minimum (not
maximum) of $2\mu$ and $2\Delta$. In addition we find another
step-like feature in the conductivity at the frequency
$\omega=2\sqrt{\mu^2+\Delta^2}$.

Thus electron-hole pairing can be detected through an observation of
 strong reflection and absorption at the frequency
$\omega=2\Delta$. Taking $d k_F\approx 0.1$, $v_F=10^8$ cm/s, and
$d=10$ nm we evaluate $\mu\approx 6$ meV and $\nu\sim\mu/h\approx
1.6$ THz. This estimate shows that for $\Delta\sim \mu$ the features
we have described can be observed in the terahertz spectral range.

\section{Surface plasmon-polaritons in the system with pairing}
\label{s5}

Surface plasmon-polariton modes in graphene are now the subject of
intensive study \cite{p1,p2,p3,p4,p5}. In particular, the interest
to graphene is connected with the possibility of modifying the
energy spectrum by external gates. The latter effect can be utilized
in the plasmonic transformation optics\cite{p6}. There are two kinds
of surface plasmon-polariton waves in graphene, the TM\cite{p7,p8}
and TE\cite{p9} ones. In the double layer system with the interlayer
distance $d$ smaller than the inverse Fermi wave number two TM
modes, the symmetric (optical) and antisymmetric (acoustic)
ones\cite{p10}, and one (symmetric) TE mode can propagate.  The
question we consider is how the pairing influences the spectrum and
damping of these modes.

\subsection{TM modes}

We consider a double layer graphene system  in a vacuum. The
graphene layers 1 and 2 are located in the  $z=+d/2$ and $z=-d/2$
planes and separated by a spacer with the dielectric constant
$\varepsilon_d$. The wave vector of the plasmon mode $\mathbf{q}$ is
directed along the
 $x$ axis.
 The electric field of the surface TM mode has the longitudinal
 ($x$)  as well as the transverse ($z$) component,
 and the magnetic field has only the transverse ($y$) component.
The electric and magnetic fields are exponentially decaying in both
directions away from the graphene layers.

 The
presence of graphene layers is taken into account by the boundary conditions
\begin{eqnarray}\label{26}
  H_{y}\big|_{z=\frac{d}{2}+0}-H_{y}\big|_{z=\frac{d}{2}-0} &=&
  -\frac{4\pi}{c} \left(\sigma_{xx}^{11} E_x\big|_{z=\frac{d}{2}}
  + \sigma_{xx}^{12} E_x\big|_{z=-\frac{d}{2}}\right),\cr
  H_{y}\big|_{z=-\frac{d}{2}+0}-H_{y}\big|_{z=-\frac{d}{2}-0}  &=&
  -\frac{4\pi}{c} \left(\sigma_{xx}^{22} E_x\big|_{z=-\frac{d}{2}}
  + \sigma_{xx}^{21} E_x\big|_{z=\frac{d}{2}}\right),
\end{eqnarray}
where
$\sigma_{xx}^{11}=\sigma_{xx}^{22}=(\sigma_{+,xx}+\sigma_{-,xx})/2$
and
$\sigma_{xx}^{12}=\sigma_{xx}^{21}=(\sigma_{+,xx}-\sigma_{-,xx})/2$
are the intralayer and interlayer  conductivity, and
$\sigma_{\pm,xx}$ is the parallel current (counterflow)
conductivity. The latter quantities are expressed through the
response functions $\Pi_{\pm,00}$:
\begin{equation}\label{28-1}
  \sigma_{\pm,xx}(q\mathbf{i}_x,\omega)= ie^2\frac{\omega}{q^2}\Pi_{\pm,00}(\mathbf{q}
,\omega).
\end{equation}

The solution of Maxwell equations with the boundary conditions
(\ref{26}) yields the dispersion equations for the symmetric
($E_x\big|_{z=\frac{d}{2}}=E_x\big|_{z=-\frac{d}{2}}$) and
antisymmetric
($E_x\big|_{z=\frac{d}{2}}=-E_x\big|_{z=-\frac{d}{2}}$) TM modes:
\begin{equation}\label{27}
    1+\frac{4\pi i
    \sigma_{+,xx}(q\mathbf{i}_x,\omega)\kappa_1}{\omega}+\frac{\varepsilon_d
    \kappa_1}{\kappa_2}\tanh\frac{\kappa_2 d}{2}=0,
\end{equation}

\begin{equation}\label{28}
    \left(1+\frac{4\pi i
    \sigma_{-,xx}(q\mathbf{i}_x,\omega)\kappa_1}{\omega}\right)\tanh\frac{\kappa_2 d}{2} +\frac{\varepsilon_d
    \kappa_1}{\kappa_2}=0,
\end{equation}
where $\kappa_1=\sqrt{q^2-\omega^2/c^2}$ and
$\kappa_2=\sqrt{q^2-\epsilon_d\omega^2/c^2}$.

In a wide range of frequencies the wave vector of the surface TM
mode is much larger than the wave vector of an electromagnetic wave
with the same frequency  in a free space ($q\gg \omega/c$). This
condition is violated only for the symmetric mode at very small $q$.
This range of $q$ is not considered here. Implying also that $q\gg
\varepsilon_d\omega/c$ we replace $\kappa_1$ and $\kappa_2$ in the
dispersion equations (\ref{27}), (\ref{28}) with $q$ and reduce
 Eqs. (\ref{27}), (\ref{28}) to the form
\begin{equation}\label{29}
    \varepsilon_{\pm}({q},\omega)=0,
\end{equation}
where
\begin{equation}\label{29a}
\varepsilon_{\pm}({q},\omega)=1+\frac{4\pi i q}{\omega}\frac{1\pm
e^{-qd}}{\varepsilon_d+1\pm(\varepsilon_d-1)e^{-qd}}\sigma_{+,xx}(q\mathbf{i}_x,\omega)
\end{equation}
are the two-dimensional(2D) dielectric functions. The functions
$\varepsilon_{\pm}$ determine the screening of the scalar potentials
$\varphi_{\pm}=\varphi_1\pm\varphi_2$ of the test charges $q_1$ and
$q_2=\pm q_1$ located in the graphene layers 1 and 2 one above the
other. The functions (\ref{29a}) can be presented in the form
\begin{equation}\label{29b}
\varepsilon_{\pm}({q},\omega)=1-V_{\pm}(q){\Pi_{\pm,00}(\mathbf{q}
,\omega)},
\end{equation}
where
\begin{equation}\label{30}
{V}_{\pm}(q)=\frac{4\pi e^2}{q}\frac{1\pm
e^{-qd}}{\varepsilon_d+1\pm(\varepsilon_d - 1)e^{-qd}}=V_{S}(q)\pm
V_{D}(q).
\end{equation}
 One can see that Eq. (\ref{29b}) corresponds to the
random phase approximation for the dielectric functions.

Eq. (\ref{29}) determines the dispersion of 2D plasmons. The
difference between 2D plasmons and three-dimensional (3D) plasmons
is the following. 3D plasmons are the longitudinal excitations of
the electric field. Plasmons in 2D conductors in a 3D space are TM
waves. The electric field of that wave has the longitudinal as well
as the transverse component. The magnetic field is also nonzero but
small in the limit $q\gg \omega/c$.

We specify the case $d k_F\ll 1$ and consider the range of wave
vectors $q<k_F$. Expanding  ${V}_{\pm}(q)$ in the small parameter
$qd$ and neglecting the higher order terms we obtain
\begin{equation}\label{31}
\varepsilon_{+}({q},\omega)=1+\frac{4\pi  i
q\sigma_{+,xx}(q\mathbf{i}_x,\omega) }{\omega},
\end{equation}
\begin{equation}\label{32}
\varepsilon_{-}({q},\omega)=1+\frac{2\pi i q^2
d\sigma_{-,xx}(q\mathbf{i}_x,\omega) }{\varepsilon_d\omega}.
\end{equation}

Let us first analyze the dispersion equation for the symmetric TM
wave. According to Eqs. (\ref{29}) and (\ref{31}) this mode can
propagate in the frequency range where
$\mathrm{Im}\left[\sigma_{+,xx}(q\mathbf{i}_x,\omega)\right]>0$.
 The qualitative analysis
can be done by replacing $\sigma_{+,xx}(q\mathbf{i}_x,\omega)$ with
$\sigma_{+}(\omega)$. Then from Fig. \ref{fig1} we see that in the
system with the pairing at $T\ne 0$ the dispersion equation
$\varepsilon_{+}({q},\omega)=0$
 may have two solutions, one is in the range
$\omega<2\Delta$, and the other, in the range $\omega>2\Delta$. The
real part of $\sigma_{+,xx}$ determines Landau damping. Since
$\mathrm{Re}\left[\sigma_+(\omega)\right]$ is extremely small at
$\omega<2\Delta$ one can expect that the low frequency solution
corresponds to weakly damped plasmons. For $\omega>2\Delta$ the real
part of conductivity is rather large and the high frequency solution
will correspond to strongly damped plasmons.

The influence of pairing on Landau damping can be understood from
the spectrum  of electron-hole single particle excitations. In the
system with pairing  all states with negative energies ($E
=-E_{\mathbf{k},\lambda}$) are filled at $T=0$, and all states with
positive energies ($E =+E_{\mathbf{k},\lambda}$) are empty.
Therefore, the continuum of  electron-hole excitations is determined
by the inequality
\begin{equation}\label{303}
E_{e-h}(q)\geq
\min[E_{\mathbf{k}+\mathbf{q},\lambda}+E_{\mathbf{k},\lambda'}],
\end{equation}
where the minimum is taken over all $\mathbf{k}$, $\lambda$ and
$\lambda'$. It follows from this inequality that $E_{e-h}(q)\geq
2\Delta$. In the normal system  the electron-hole excitation
continuum starts from $E_{e-h}(q)=0$. Fig. \ref{fig3} demonstrates
the modification of the electron-hole excitation continuum under the
pairing. One can see  that in the paired state Landau damping for
the modes with energies $\omega(q)<2\Delta$ is suppressed, while the
modes with $\omega(q)>2\Delta$ may suffer from a strong damping.

\begin{figure}
\begin{center}
\includegraphics[width=8cm]{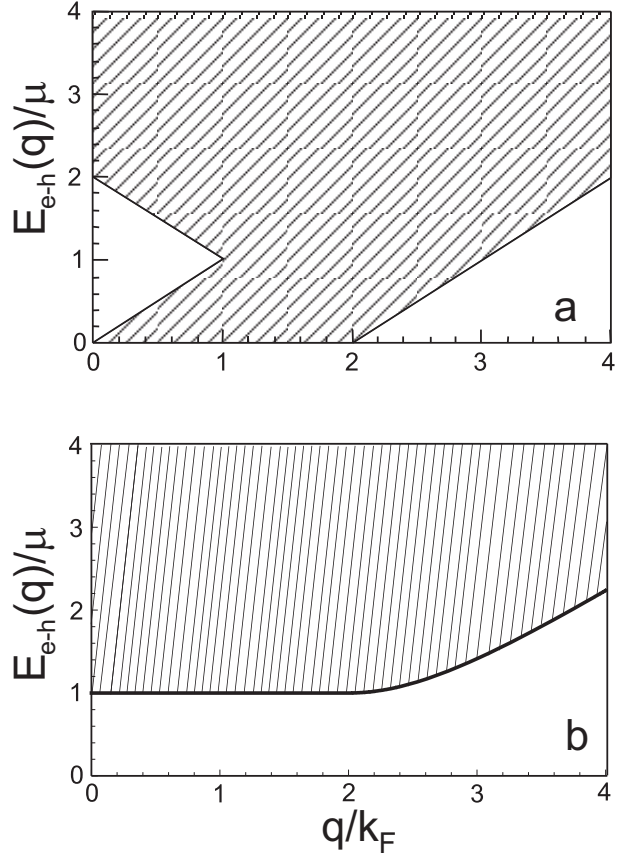}
\end{center}
\caption{The continuum of  electron-hole  excitations (shaded area)
for the double layer graphene system  in the normal state (a) and in
the paired state with $\Delta=0.5\mu$ (b). } \label{fig3}
\end{figure}

We compute the spectrum of the symmetric TM wave from the equation
\begin{equation}\label{304}
    \mathrm{Re}[\varepsilon_+({q},\omega)]=0,
\end{equation}
where $\varepsilon_+({q},\omega)$ is given by Eq. (\ref{31}) in
which the dependence of $\sigma_{+,xx}$ on $q$ is taken into account. The
damping rate is evaluated as
\begin{equation}\label{305}
    \gamma(q)=\frac{\mathrm{Im}[\varepsilon_+({q},\omega(q))]} {\frac{\partial
\mathrm{Re}[\varepsilon_+({q},\omega)]}{\partial
\omega}\Big|_{\omega=\omega(q)}},
\end{equation}
where $\omega(q)$ is the solution of Eq. (\ref{304}).  Since Eqs.
(\ref{304}), (\ref{305}) are valid at small damping we do not
consider the solutions of  Eq. (\ref{304}) with $\gamma(q)\gtrsim
\omega(q)$.

The results of computation for $\Delta=0.5\mu$ at $T=0.1\mu$ and
$T=0.2\mu$ are presented in Fig. \ref{fig4}. One can see that in the
state with the pairing the symmetric TM mode splits into two
branches (the spectrum in the normal state is also shown in Fig.
\ref{fig4}). In the long wavelength range the lower branch is a
weakly damped one. There is a critical wave vector $q_c$ above which
the solution of Eq. (\ref{304}) that corresponds to the lower branch
disappears. At $q$ approaching $q_c$ the damping rate for the lower
branch increases sharply.  Under increase in temperature the
frequency of this mode and the critical wave vector $q_c$ grows up .
The lower branch is a thermally activated mode. At $T=0$ this mode
does not exist. It can be seen from the explicit expression for the
response function $\Pi_{+,00}(\mathbf{q},\omega)$ (Eq. (\ref{22a})).
At $T=0$ this response function is real and negative at
$\omega<2\Delta$. Therefore
$\mathrm{Im}\left[\sigma_{+,xx}(\mathbf{q},\omega)\right]<0$ and
 Eq. (\ref{304}) has no solution in the
frequency range $\omega<2\Delta$ at zero temperature.  The behavior
of the lower mode allows considering it as one connected with
plasmon oscillations of the normal component decoupled from the
superfluid one.

The frequency of the upper branch is restricted from below by the
inequality $\omega>2\Delta$. The damping rate for this mode is much
higher than for the lower branch. This branch exists in the wave
vector range $q>q_{c1}$. At $q$ close to $q_{c1}$ the frequency of
this mode approaches  $2\Delta$ and its damping rate increases
sharply (the mode becomes overdamped).

In Fig. \ref{fig5} the spectrum and the damping rate for the lower
and upper branches at $\Delta=0.2\mu$ and $T=0.1\mu$  are shown.
Comparing Fig. \ref{fig5} and Fig. \ref{fig4} we conclude that
lowering of $\Delta$ (at the same $T$) results in an expansion of
the wave vector range for two branches and in a decrease of the
frequency and the damping rate of the upper branch.

\begin{figure}
\begin{center}
\includegraphics[width=8cm]{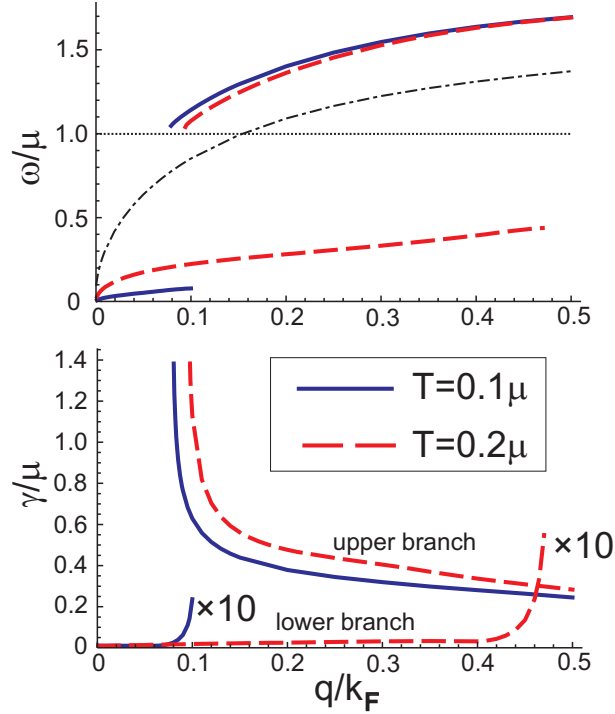}
\end{center}
\caption{The spectrum (upper panel) and damping rate (lower panel)
for two branches of the symmetric TM mode in the system with
electron-hole pairing  at $\Delta=0.5\mu$ and two different
temperatures $T=0.1\mu$ and $T=0.2\mu$. In the upper panel the
spectrum of the symmetric TM mode in the normal  state is shown by
dash-dotted line and the lower boundary for the continuum of
electron-hole  excitations is shown by dotted line. } \label{fig4}
\end{figure}

\begin{figure}
\begin{center}
\includegraphics[width=8cm]{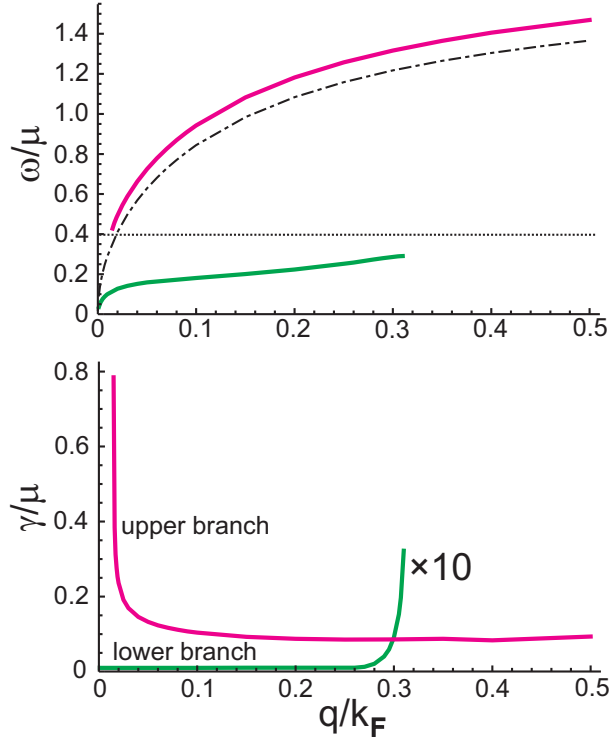}
\end{center}
\caption{The same as in Fig. \ref{fig4} at $\Delta=0.2\mu$ and
$T=0.1\mu$. } \label{fig5}
\end{figure}

The plasmon mode spectrum can also be extracted from the energy loss
function
\begin{equation}\label{306}
 L_\pm({q},\omega)=-\mathrm{Im}\left[\frac{1}{\varepsilon_{\pm}({q},\omega)}\right].
\end{equation}
This function determines the losses of energy of a pair of test
charges oscillating  with the frequency $\omega$  located in the
adjacent graphene layers opposite to each other. The losses are
connected with the excitation of plasmons at this frequency. A sharp
peak in the loss function corresponds to  a weakly damped plasmon
mode. In Fig. \ref{fig6} the dependence of $L_+({q},\omega)$ on
$\omega$ at fixed $q$ is presented for $T=0$ and $T=0.1\mu$.  At
$T=0.1\mu$  this dependence contains a sharp peak that corresponds
to the lower mode and a wide peak that corresponds to the upper
mode. The positions of the peaks depend on $\Delta$. At $T=0$ the
peak that corresponds to the lower mode disappears, while the upper
mode peak remains unchanged. There is only one peak at $\Delta=0$.
It corresponds to the symmetric TM mode in the normal state. The
behavior of the loss function demonstrates splitting of the TM wave
into the lower and upper branches. The lower one is a weakly damped
and thermally activated mode. The upper one is a strongly damped
mode, and practically unsensitive to the temperature in the
temperature range considered.

\begin{figure}
\begin{center}
\includegraphics[width=8cm]{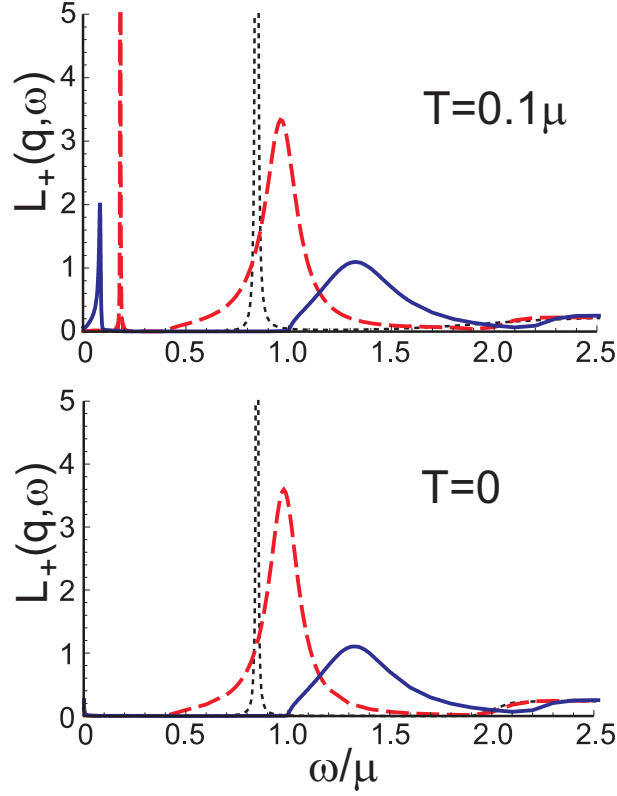}
\end{center}
\caption{Frequency dependence of the loss function $L_+({q},\omega)$
for $q=0.1k_F$ at $T=0.1\mu$ (upper panel) and $T=0$ (lower panel)
in the paired state with $\Delta=0.5\mu$ (solid line) and
$\Delta=0.2\mu$ (dashed line), and in the normal state (dotted
line). } \label{fig6}
\end{figure}

 Concluding this analysis we note that the splitting of the plasmon mode into
a number of branches may also for the system is subjected by a
strong magnetic field directed perpendicular to the graphene layers
(without electron-hole pairing)\cite{b2}.

Let us now switch to the antisymmetric mode.  The gauge-invariant
response function $\Pi_{-,00}$ is given by Eq. (\ref{7d}) with the
renormalized vertex function (\ref{18f}). The result of computation
can be presented in the form
\begin{equation}\label{330}
\Pi_{-,00}(\mathbf{q},\omega)=\Pi^{(1)}_{-,00}(\mathbf{q},\omega)
+\Pi^{(2)}_{-,00}(\mathbf{q},\omega),
\end{equation}
where the first term corresponds to the bare vertex approximation
\begin{eqnarray}\label{322a}
\Pi^{(1)}_{-,00}(\mathbf{q},\omega)=-{ 4}\sum_{\lambda,\lambda'}\int
\frac{d \mathbf{k}}{(2\pi)^2}
F^{00}_{\lambda,\lambda',\mathbf{k},\mathbf{q}}\Bigg[
\left(P^{00}_{\lambda,\lambda',\mathbf{k},\mathbf{q}}+\frac{\Delta^2}{EE'}\right)\frac{(1-f'
-f)(E+E')} {(E+E')^2-\omega^2}\cr
+\left(L^{00}_{\lambda,\lambda',\mathbf{k},\mathbf{q}}-\frac{\Delta^2}{EE'}\right)\frac{(f'-f
)(E-E')} {(E-E')^2-\omega^2}\Bigg], \end{eqnarray} and the second
term  is caused by the renormalization of the bare vertex functions
(\ref{18f})
\begin{equation}\label{331}
  \Pi^{(2)}_{-,00}(\mathbf{q},\omega)={4}\frac{\omega^2}{\omega^2 -s^2 q^2}\sum_{\lambda,\lambda'}\int
\frac{d \mathbf{k}}{(2\pi)^2}
F^{00}_{\lambda,\lambda',\mathbf{k},\mathbf{q}}\frac{\Delta^2}{E
E'}\Big[ \frac{(1-f' -f)(E+E')} {(E+E')^2-\omega^2}-\frac{(f'-f
)(E-E')} {(E-E')^2-\omega^2}\Big].
\end{equation}

 At $\omega=0$ the second term in
Eq. (\ref{330}) vanishes. Therefore, the static response function is
gauge invariant in the bare vertex approximation. The conductivity
$\sigma_{-,xx}(q\mathbf{i}_x,\omega)$ is determined by Eq.
(\ref{28-1}). In the limit $q\to 0$ this equation
 gives the uniform conductivity
$\sigma_-(\omega)$. The right hand side of Eq. (\ref{28-1}) has a
finite limit  if $\Pi_{-,00}(0,\omega)=0$. At $q\to 0$ the quantity
$\Pi^{(2)}_{-,00}(\mathbf{q},\omega)$ compensates the terms
proportional to $\Delta^2$ in $\Pi^{(1)}_{-,00}(\mathbf{q},\omega)$,
and in this limit the expression for $\Pi_{-,00}(\mathbf{q},\omega)$
(Eq. (\ref{330})) coincides with one for
$\Pi_{+,00}(\mathbf{q},\omega)$ (Eq. (\ref{22a})). Using the
explicit expression (\ref{22a}) one can show analytically that
indeed $\Pi_{-,00}(0,\omega)=\Pi_{+,00}(0,\omega)=0$. In other
words, the vertex function $\Gamma_0^-$ in the form
 Eq. (\ref{18f}) ensures the applicability of Eq. (\ref{28-1}). In the contrary,
computations with the bare vertex $\gamma_0^-$ instead of the
dressed one $\Gamma_0^-$ would yield an unphysical answer for
$\sigma_-(\omega)$.

As above, for the qualitative analysis one can neglect the
dependence of $\sigma_{-,xx}$ on $q$ in Eq. (\ref{32}). The
dependence ${\rm Im}[\sigma_-(\omega)]$ computed from Eqs.
(\ref{28-1}) and (\ref{330}) is shown in Fig. \ref{fig7}. One can
see that the imaginary part of $\sigma_-(\omega)$ remains almost
unchanged under the pairing and we expect only an inessential impact
of the pairing on the spectrum of the antisymmetric TM mode.

\begin{figure}
\begin{center}
\includegraphics[width=8cm]{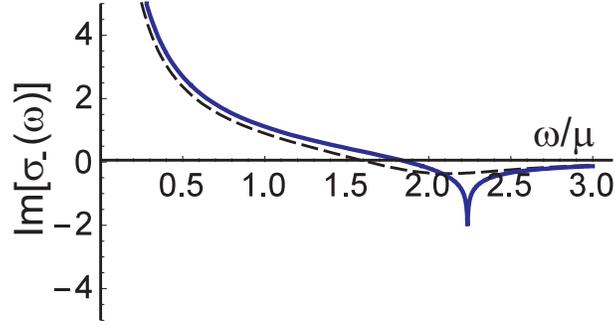}
\end{center}
\caption{The imaginary part of the counterflow conductivity (in
units of $e^2/4\hbar$) at $T=0.1\mu$ in the paired state with
$\Delta=0.5\mu$ (solid line) and in the normal state (dashed line).
} \label{fig7}
\end{figure}

The spectrum of the antisymmetric TM mode computed from the equation
$\mathrm{Re}[\varepsilon_-(q,\omega)]=0$ with
$\varepsilon_-(q,\omega)$ given by Eq. (\ref{32}) is shown in Fig.
{\ref{fig8}}. The parameters $\varepsilon_d=4$ and $k_F d=0.2$ are
used for the computations.  Fig. {\ref{fig8}} shows that the
antisymmetric TM mode remains the acoustic one in the state with the
pairing and its velocity slightly reduces comparing to one in the
normal state. In Fig. \ref{fig9} the loss function $L_-(q, \omega)$
is presented. Fig. \ref{fig9} demonstrates that the antisymmetric
mode is a weakly damped one in the normal as well as  in the paired
state. We remind that the renormalized vertex function (\ref{18f})
was obtained in the limit $\omega\ll \Delta$ and its applicability
at frequencies $\omega\gtrsim \Delta$ is questionable.

\begin{figure}
\begin{center}
\includegraphics[width=8cm]{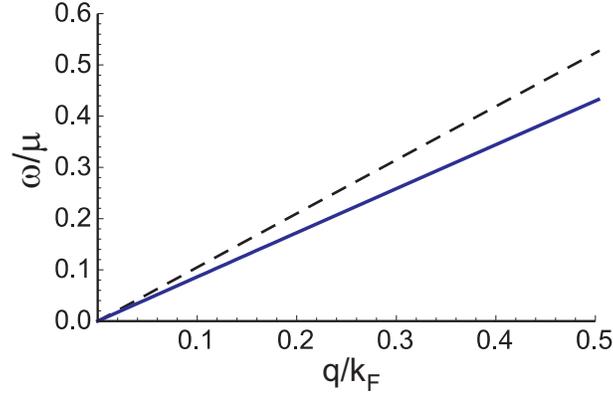}
\end{center}
\caption{The spectrum of the antisymmetric TM wave at $T=0.1\mu$ in
the paired state with $\Delta=0.5\mu$ (solid line) and in the normal
state (dashed line).} \label{fig8}
\end{figure}

The fact that the pairing does not influence the antisymmetric TM
mode can be understood as follows. This mode corresponds to
out-of-phase oscillations of electron densities in graphene layers.
Out-of-phase oscillations in electron densities are equivalent to
in-phase oscillations in electron and hole densities. It is quite
natural that the electron-hole pairing does not suppress such
oscillations.

\begin{figure}
\begin{center}
\includegraphics[width=8cm]{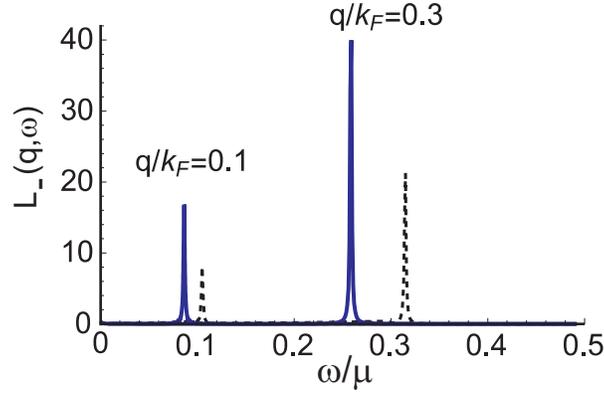}
\end{center}
\caption{Frequency dependence of the loss function $L_-({q},\omega)$
for $q=0.1k_F$ and $q=0.3k_F$ at $T=0.1\mu$ in the paired state with
$\Delta=0.5\mu$ (solid line)  and in the normal state (dotted
line).} \label{fig9}
\end{figure}

\subsection{TE modes}

Surface TE  waves with the wave vector $\mathbf{q}$ directed along
the $x$ axis have the transverse electric component $E_y$ and the
longitudinal $H_x$ as well as the transverse $H_z$ magnetic
component. The boundary conditions for the tangential component of
the magnetic field are
\begin{eqnarray}\label{26a}
  H_{x}\big|_{z=\frac{d}{2}+0}-H_{x}\big|_{z=\frac{d}{2}-0} &=&
  \frac{4\pi}{c} \left(\sigma_{yy}^{11} E_y\big|_{z=\frac{d}{2}}
  + \sigma_{yy}^{12} E_y\big|_{z=-\frac{d}{2}}\right),\cr
  H_{x}\big|_{z=-\frac{d}{2}+0}-H_{x}\big|_{z=-\frac{d}{2}-0}  &=&
  \frac{4\pi}{c} \left(\sigma_{yy}^{22} E_y\big|_{z=-\frac{d}{2}}
  + \sigma_{yy}^{21} E_y\big|_{z=\frac{d}{2}}\right).
\end{eqnarray}
Maxwell equations with the boundary conditions (\ref{26a}) yield the
dispersion equations for the TE waves:
\begin{equation}\label{411}
    1-\frac{k}{\kappa}\tan\frac{k d}{2}-\frac{4\pi i \omega
    \sigma_{+,yy}(q\mathbf{i}_x,\omega)}{c^2 \kappa}=0,
\end{equation}
\begin{equation}\label{412}
    1+\frac{k}{\kappa}\cot\frac{k d}{2}-\frac{4\pi i \omega
    \sigma_{-,yy}(q\mathbf{i}_x,\omega)}{c^2 \kappa}=0,
\end{equation}
where $\kappa=\sqrt{q^2-\omega^2/c^2}$ and
$k=\sqrt{\varepsilon_d\omega^2/c^2-q^2}$. Eq. (\ref{411})
corresponds to the symmetric TE wave, and Eq. (\ref{412}), to the
antisymmetric TE wave.

At $\sigma_{\pm} =0$  (no graphene layers) and the general  $d$ Eqs.
(\ref{411}), (\ref{412}) have a number of solutions which correspond
to the symmetric and antisymmetric waveguide modes.  At $q d\lesssim
1$ only the lowest symmetric mode survives. The presence of graphene
layers influences only the modes with the phase velocities close to
the light velocity $c$. We restrict our analysis with frequencies
comparable or smaller than the chemical potential $\mu$. The mode
with the phase velocity $v\approx c$ and the frequency $\omega
\lesssim \mu$ has the wave vector $q\lesssim k_F v_F/c\ll k_F$.
Since the pairing occurs at $k_F d<1$, the strong inequality $qd\ll
1$ is fulfilled. In this limit the real part of Eq. (\ref{411}) is reduced to
\begin{equation}\label{411a}
    \sqrt{1-\frac{v^2}{c^2}}=\pi\alpha \left[C\tilde{\omega} - \mathrm{Im}(\tilde{\sigma}_+(\omega))\right],
\end{equation}
where  $\alpha\approx 1/137$ is the fine structure constant,
$\tilde{\sigma}_+(\omega)$ is the conductivity normalized to
$e^2/4\hbar$, $\tilde{\omega}=\omega/\mu$ and
$C=dk_F(\varepsilon_d-1)/2\pi\alpha_{eff}$ is the material
parameter. In  Eq. (\ref{411a}) we take into account that $q$ is in
3 orders smaller than $k_F$ and replace  the finite wave vector
conductivity with the uniform conductivity $\sigma_+(\omega)$. Eq.
(\ref{411a}) determines the phase velocity $v$ as the function of
$\omega$. Eq. (\ref{411a}) has a solution  in the frequency range,
where  $\mathrm{Im}(\tilde{\sigma}_+(\omega))<C \tilde{\omega}$. For
typical parameters ($d k_F<0.2$, $\alpha_{eff}=2.2$, and
$\varepsilon_d=4$) the constant $C$ is very small ($C< 0.05$) and it
can be neglected. In fact, the TE wave frequency range is determined
by the same condition as one for the monolayer graphene
 ($\mathrm{Im}(\tilde{\sigma}_+(\omega))<0$).

In the normal state the TE mode exists only at $\omega>1.67 \mu$.
One can see from Fig. \ref{fig1}b that  the pairing opens  the low
frequency window $\omega_{min}<\omega<2\Delta$ for the TE mode. At
temperature $T\to 0$  the condition
$\mathrm{Im}(\tilde{\sigma}_+(\omega))<0$ is fulfilled for
$0<\omega<2\Delta$ (see Fig. \ref{fig10}) and  the lower edge
$\omega_{min}$ goes to zero.

\begin{figure}
\begin{center}
\includegraphics[width=8cm]{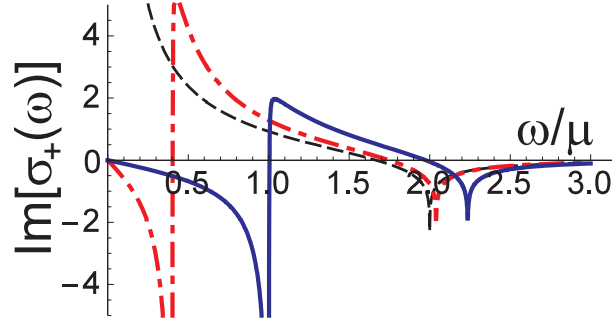}
\end{center}
\caption{The imaginary part of the parallel current conductivity (in
$e^2/4\hbar$ units) at $T=0$ in the paired state with
$\Delta=0.5\mu$ (solid line) and $\Delta=0.2\mu$ (dash-dotted line),
and in the normal state (dashed line). } \label{fig10}
\end{figure}

\section{Conclusion}

In conclusion, we have shown that  the electron-hole pairing
significantly changes spectral properties of double layer graphene
systems in the terahertz range. The pairing causes the appearance of
sharp high peaks in the absorption and reflection at the frequency
$\omega=2\Delta$ and a rather large (much larger than in an undoped
graphene) absorption at $\omega>2\Delta$.

The pairing influences essentially the surface symmetric TM mode.
This mode splits into the lower and upper branches. The lower branch
has the frequency $\omega<2\Delta$. It is practically undamped and
appears only at nonzero temperatures in the long wavelength range.
The spectrum of the lower branch is strongly temperature dependent.
The upper branch frequency is in the range $\omega > 2\Delta$, and
this mode is strongly damped.

The influence of pairing on the antisymmetric TM mode is
inessential.

It is established that in the paired state  a low frequency TE mode
can propagate. In the normal state the frequency range for such a
mode is restricted from below by the inequality $\omega>1.67 \mu$.
In the paired system the additional frequency window
$\omega_{min}<\omega<2\Delta$ opens, where $\omega_{min}$ goes to
zero at $T\to 0$.

Other Dirac double layer systems such as thin topological insulator
plates, double layer silicene, germanene and $\alpha$-graphyne
structures will demonstrate the same behavior. We also expect
qualitatively the same features in double layer electron-hole
systems made of a pair of bilayer\cite{bil} or few-layer graphene
\cite{mult} sheets.

The appearance of additional peaks in the transmission, reflection,
and absorption spectra of double layer graphene systems would be a
hallmark of the electron-hole pairing. A strong modification of the
spectrum of an optical TM mode with temperature can be used in
plasmonics for creating transformation optic devices with a thermal
control.

\end{document}